\documentclass[twocolumn,runningheads]{svjour2}
\smartqed  
\usepackage{graphicx}
\usepackage{mathptmx}   
\usepackage{amsmath}
\usepackage{amsfonts}
\usepackage{amssymb}
\usepackage{latexsym}
\usepackage[authoryear]{natbib}
\usepackage{journals}

\journalname{Astrophysics and Space Science (CoRoT/ESTA Volume)}
\newcommand{\rsol}{\mbox{${\mathrm R}_{\odot}$}}
\newcommand{\msol}{\mbox{${M}_{\odot}$}}

\newcommand{\corot}{{\small CoRoT}}

\newcommand{\esta}{{\small ESTA}}

\newcommand{\task}{{\small\sc TASK}}
\newcommand{\tasks}{{\small\sc TASKs}}

\newcommand{\ASTEC}{{\small ASTEC}}
\newcommand{\ATON}{{\small ATON}}
\newcommand{\CESAM}{{\small CESAM}}
\newcommand{\CLES}{{\small CL\'ES}}
\newcommand{\FRANEC}{{\small FRANEC}}
\newcommand{\TGEC}{{\small TGEC}}
\newcommand{\GARSTEC}{{\small GARSTEC}}
\newcommand{\GENEC}{{\small GENEC}}
\newcommand{\STAROX}{{\small STAROX}}
\newcommand{\YREC}{{\small YREC}}
\newcommand{\ADIPLS}{{\small ADIPLS}}
\newcommand{\FILOU}{{\small FILOU}}
\newcommand{\GRACO}{{\small GRACO}}
\newcommand{\LOSC}{{\small LOSC}}
\newcommand{\NOSC}{{\small NOSC}}
\newcommand{\OSCROX}{{\small OSCROX}}
\newcommand{\POSC}{{\small POSC}}
\newcommand{\ROMOSC}{{\small LNAWENR}}
\newcommand{\PULSE}{{\small PULSE}}

\newcommand{\astec}{{\small ASTEC}}
\newcommand{\cesam}{{\small CESAM}}
\newcommand{\cles}{{\small CL\'ES}}
\newcommand{\franec}{{\small FRANEC}}
\newcommand{\tgec}{{\small TGEC}}
\newcommand{\garstec}{{\small GARSTEC}}

\def\elem#1[#2]{{}$^{\scriptsize #2}${\rm #1}}

\newcommand{\PMS}{{\small PMS}}
\newcommand{\MS}{{\small MS}}
\newcommand{\ZAMS}{{\small ZAMS}}
\newcommand{\TAMS}{{\small TAMS}}
\newcommand{\SGB}{{\small SGB}}

\begin{document}

\title{The CoRoT Evolution and Seismic Tools Activity
}
\subtitle{Goals and Tasks}

\titlerunning{Goals and Tasks of ESTA-CoRoT}  

\author{Y.~Lebreton \and
        M.~J.~P.~F.~G.~Monteiro \and
        J.~Montalb\'an \and
	A.~Moya \and
        A.~Baglin \and
	J.~Christensen-Dalsgaard \and
	M.-J. Goupil \and
	E.~Michel \and
	J. Provost \and
	I.W. Roxburgh \and
	R. Scuflaire and the ESTA Team
}


\institute{Y.~Lebreton \at
           Observatoire de Paris, GEPI, CNRS UMR 8111,
           5 Place Janssen, 92195 Meudon, France \\
           \email{Yveline.Lebreton@obspm.fr}
       \and
           M.J.P.F.G.~Monteiro \at
           Centro de Astrof\'{\i}sica da Universidade do Porto and 
           Departamento de Matem\'atica Aplicada da Faculdade de Ci\^encias,
           Universidade do Porto, Portugal
       \and
	      J.~Montalb\'an \and R.~Scuflaire
	      \at Institut d'Astrophysique et Geophysique,
	      Universit\'e de Li\`ege, Li\`ege, Belgium
	\and
	      A.~Moya 
	      \at Instituto de Astrof\'isica de Andaluc\'ia- CSIC,
	      Granada, Spain
	\and
	      A.~Baglin \and M.-J. Goupil \and E.~Michel \and I.W.~Roxburgh
	      \at Observatoire de Paris, LESIA, France
	\and
	      J.~Christensen-Dalsgaard 
	      \at Institut for Fysik og Astronomi, Aarhus Universitet, Aarhus, Denmark
	\and
	      J.~Provost
	      \at Cassiop\'ee, URA CNRS 1362, Observatoire de la C\^ote d'Azur, Nice, France
	\and
	      I.W.~Roxburgh
	      \at Queen Mary University of London, England
}

\date{Received: date / Accepted: date}

\maketitle
\begin{abstract}

The forthcoming data expected from space missions such as CoRoT require the capacity of the available tools to provide accurate models whose numerical precision is well above the expected observational errors.
In order to secure that these tools meet the specifications, a team has been established to test and, when necessary, to improve the codes available in the community.
The CoRoT evolution and seismic tool activity (ESTA) has been set up with this mission.

Several groups have been involved. The present paper describes the motivation and the organisation of this activity,
providing the context and the basis for the presentation of the results that have been achieved so far.
This is not a finished task as future even better data will continue to demand more precise and complete tools for asteroseismology. 

\keywords{ stars: interiors \and stars: evolution \and stars: oscillations \and methods: numerical}
\PACS{97.10.Cv \and 97.10.Sj \and 95.75.Pq}
\end{abstract}

\section{Introduction}\label{sec:intro}

The {\corot} satellite was launched into space on December 27, 2006. This experiment will provide us with stellar oscillation data (frequencies, amplitudes, line widths) for stars of various masses and chemical compositions -- mainly main sequence solar type stars,  $\delta$ Scuti stars  and  $\beta$ Cephei stars -- with an expected accuracy on the frequencies of a few $10^{-7}\ {\mathrm{Hz}}$ \citep{2002sshp.conf...17B,michel06}. 

Such an accuracy is needed to determine some of the key features of the stellar interiors such as the extent of the convective core of intermediate mass stars \citep{1999ASPC..173..257R,2006MNRAS.372..949M,2007ApJ...666..413C} or the position of the external convection zone \citep{1994A&A...282...73A,2000MNRAS.316..165M,2004A&A...423.1051B,2006ApJ...638..440V} and the envelope helium abundance in low mass solar type stars \citep{1998IAUS..185..317M,2004MNRAS.350..277B,2006ApJ...638..440V}. High-quality seismic data also make possible inverse analyses
to infer the stellar density  \citep{1993MNRAS.264..522G} 
and rotation profiles from separated multiplets \citep{1996A&A...305..487G}.
More generally, it is expected that {\corot} will provide constraints on those aspects of the physics of stellar interiors which are still poorly understood, in particular on convection \citep{2005JApA...26..171S} and on transport mechanisms at work in the radiative zones such as microscopic diffusion and rotationally induced transport \citep{2004sshp.conf..133G,2005A&A...437..553T}. As a result stellar models will be improved by seismological inferences, and if they are combined with high-quality observational data of the global stellar parameters (luminosity, effective temperature, radius, abundances), it will allow to improve the determination of those stellar parameters not directly accessible through observation like ages and masses of single stars with valuable returns on the understanding of galactic structure and evolution \citep{2005tdug.conf..493L,2002sshp.conf..291M}.

However, studies in helioseismology have taught us that in order to probe fine details of the solar internal structure, we need both the constraints of high-quality seismic data and extremely  accurate numerical solar models \citep{1995MNRAS.274..899R,1996Sci...272.1296G,2002sshp.conf..291M}. 
Therefore, to be able to draw valuable information from the future {\corot} data, we have to ensure that we will be able to interpret them with models having reached the optimal level of accuracy.

In the 1990's the {\small GONG} solar model comparison project has been one of the first attempts to compare thoroughly solar structure models calculated with stellar evolution codes differing both in the numerical procedures and input physics. Extensive comparisons were made which allowed to better understand differences between models and to correct or improve the codes under comparison \citep{JCD88,1991LNP38851G,1998ESASP.418..555T}. 

In the same spirit, within the {\corot} Seismology Working Group, the {\esta} group has been set up (see Sect. \ref{sec:esta}), with the aim to extensively test, compare and optimise the numerical tools used to calculate stellar models and their oscillation frequencies. The goals of the {\esta} group have been (i) to be able to produce theoretical seismic predictions by means of different numerical codes and to understand the possible differences between them and (ii) to bring stellar models at the level of accuracy required to interpret the future {\corot} seismic data.

In this introductory paper, we outline the tools and specifications of the comparisons that will be presented in detail in the different papers in this volume.  Sect.~\ref{sec:corot-stars} provides a brief overview of the characteristics of the {\corot} seismology targets to be modeled. Sect. \ref{sec:esta} presents the {\esta} group (participants, tools, past meetings and publications) and briefly introduces the numerical tools (stellar evolution codes and seismic codes) used in the successive steps of the comparison work. In Sect. \ref{sec:tasks}, we present the three {\esta} tasks we have focused on. In Sect. \ref{sec:physics} we give the specifications of the tasks  concerning the input physics and parameters of the stellar models. Finally in Sect. \ref{sec:grids} we present reference grids of stellar models and associated oscillation frequencies calculated by {\esta} participants in parallel to the main {\esta} work. These grids have been used all along the preparation of the {\corot} mission and have been made available on the {\esta} web site.

\section{The observational seismic program of {\corot}}
\label{sec:corot-stars}

\begin{figure}
\centering
\resizebox{\hsize}{!}{\includegraphics{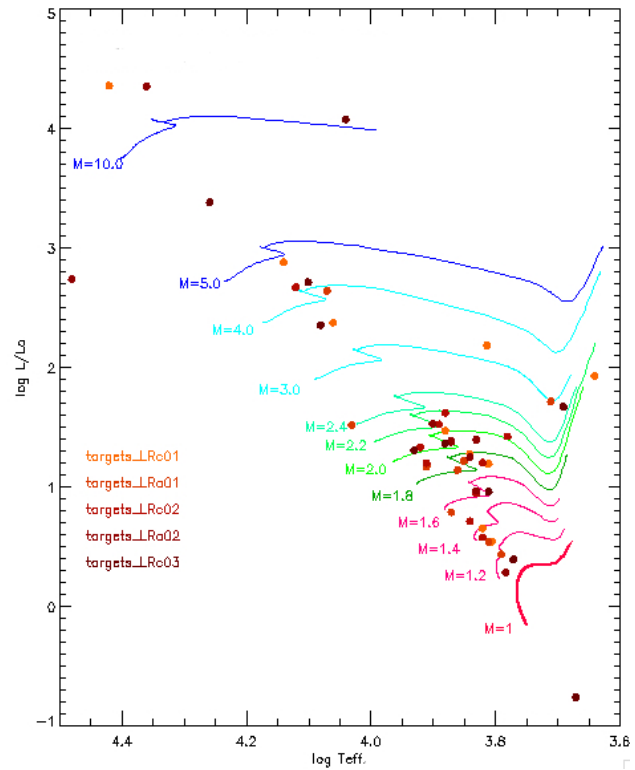}}
\caption{H--R diagram of the targets of the 5 first long runs of the {\corot} mission, as planned presently.
Long Runs are numbered as LRx0Y where x is {\em a} or {\em c} depending on whether it is a direct run in the center or anticenter direction and Y is the order ot the run in the given direction. Standard evolutionary tracks, labelled with their mass (in solar units) and corresponding to solar metallicity, no overshooting, are also shown \citep[see][]{yl2-apss}.
 }
\label{fig:targets}       
\end{figure}


The {\corot} mission has been designed and built to satisfy the scientific specifications as expressed in the document COR-SP-0-ETAN-83-PROJ, written in 2001.
At that time more recent developments of the seismology theory as well as discoveries of pulsations were evidently not taken into account.
More precisely \corot\ has been required to reach:
\begin{itemize}
\item [$\bullet$] a photometric noise of 0.6~ppm in five days in the 0.1 to 10~mHz frequency domain,
\item [$\bullet$] a duty cycle of 90\%,
\item [$\bullet$] a percentage of polluted frequencies less than 5\% of the frequency interval.
\end{itemize}

The major goals of the seismology programme were then to detect solar type oscillations in a few bright objects, covering the H--R diagram and to see whether classical pulsators ($\delta$ Scuti and $\beta$ Cephei stars in particular) near the main sequence had also lower amplitude modes.
With these goals in mind the {\em seismology program} has been defined with two components:
\begin{itemize}
\item [$\star$] the central program having at least 5 sessions of 150 days (see Fig.~\ref{fig:targets}),
\item [$\star$] the exploratory program with a few short sessions of 20 days, to complement the H--R diagram coverage.
\end{itemize}

The instrument has two different fields:
\begin{itemize}
\item [$\bullet$] the seismology field in which 10 bright stars are observed at the same time ($5.5 {<} m_V {<} 9.5$) with a sampling of 32 seconds,
\item [$\bullet$] the exoplanet field which observes 12 000 stars at the same time ($11 {<} m_V {<} 16$) with a sampling of 512 seconds and in a few cases 32 seconds. In this field the required photometric accuracy is $7{\times} 10^{{-}4}$ in 1 hour for a 15.5 magnitude star.
\end{itemize}

After one year of operations the mission fulfils all its specifications.

\section{ESTA organisation and tools}\label{sec:esta}

The Evolution and Seismic Tools Activity (ESTA) has been set up in 2002, following the ESA announcement of opportunity for participation in the CoRoT mission. The actual work has started in 2004 at the kick-off meeting that took place during the CoRoT Week 7 in Granada (December 2004).
Following this several other meetings and workshops (7 so far) have been organised as the activities progressed.
A website has been created to coordinate the activities, allowing for an efficient exchange of data and documentation as well as to share the results of the tasks that have been organised around the key objectives of ESTA. This site is available at the following url:
$$
{\mathtt {http://www.astro.up.pt/corot/}}
$$

ESTA has been open to participation of the whole community and in particular the involvement of groups with evolution codes or seismic codes has been actively pursued.
Up to now there are over 40 team members involved in one or several of the tasks corresponding to more than 16 institutes/groups spread over 9 countries. 
It includes the participation of 10 evolution codes and 9 oscillation codes in the tasks and comparisons that have been developed over the last three years.
As part of the effort several results have been published (as papers or proceedings of the workshops) or made available through the web-page.

\subsection{Stellar internal structure and evolution codes}\label{sec:code-evo}

Ten stellar evolution codes have participated in \esta\ at different levels. The list is given below. Each of them is presented in detail in a specific paper in this volume.

\begin{itemize}

\item [$\bullet$] \ASTEC\ -- 
The {\em Aarhus Stellar Evolution Code} is a Danish stellar evolution code that has been involved in all the \esta\ \tasks. It is presented by \citet{jcd1-apss}.
  
\item [$\bullet$] \ATON\ -- This is an Italian stellar evolution code. Up to now it has participated only marginally in \esta\ activities. It is presented by \citet{pv-apss}. 

\item [$\bullet$] \CESAM\ -- The {\em Code d'\'Evolution Stellaire Adaptatif et Modulaire} is a French stellar evolution code. This public code\footnote{available at {\tt\small http://www.obs-nice.fr/cesam}} has been involved in all the \esta\ \tasks\ as well as in the reference grids calculation. It is presented by \citet{pm-apss}. 
  
\item [$\bullet$] \CLES\ -- The {\em Code Li\'egeois d'\'Evolution Stellaire} is a Belgian stellar evolution code that has been involved in all the \esta\ \tasks\ as well as in the reference grids calculation. It is presented by \citet{rs1-apss}. 
  
\item [$\bullet$] \FRANEC\ -- The {\em Frascati Raphson Newton Evolutionary Code} is an Italian stellar evolution code that has participated in \esta\ \tasks~1 and 3. It is presented by \citet{sd-apss}.

\item [$\bullet$] \GARSTEC  -- The {\em Garching Stellar Evolution Code} is a German stellar evolution code that has participated in \esta\ \tasks~1 and 3. It is presented by \citet{aw-apss}.

\item [$\bullet$] \GENEC\ -- The {\em Geneva Stellar Evolution Code} is a Swiss stellar evolution code that has participated in \esta\ \task~1. It is presented by \citet{pe-apss}.

\item [$\bullet$] \STAROX\ -- This is an English stellar evolution code that has participated in \esta\ \task~1. It is presented by \citet{ir1-apss}.
  
\item [$\bullet$] \TGEC\ -- The {\em Toulouse-Geneva Evolution Code} is a French stellar evolution code that has participated in \esta\ \tasks~1 and 3. It is presented by \citet{ah-apss}.

\item [$\bullet$] \YREC\ -- The {\em Yale Rotating Stellar Evolution Code} is a US stellar evolution code. Up to now it has participated only marginally in \esta\ activities. It is presented by \citet{cs-apss}. 

\end{itemize}

\subsection{Stellar oscillation codes}\label{sec:code-osc}

Nine stellar oscillation codes have been used for the comparisons of {\esta} {\task}~2. The list is given below. Each of them is presented in details in a specific paper in this volume.

\begin{itemize}

\item [$\bullet$] \ADIPLS\ -- The {\em Aarhus Adiabatic Pulsation Package}\footnote{available at {\tt\small http://astro.phys.au.dk/$\sim$jcd/adipack.n}} contains the \ADIPLS\ stellar oscillation code itself but also many programs to manipulate the stellar model files and the output of the pulsation program (cf. Sect.~\ref{sec:physics}). It is presented by \citet{jcd2-apss}.
  
\item [$\bullet$] \FILOU\ -- This stellar oscillations code developed at Paris-Meudon Observatory is presented by \citet{jcs-apss}.

\item [$\bullet$] \GRACO\ -- The {\em Granada Oscillation Code} is presented by \citet{am-apss}.

\item [$\bullet$] \LOSC\ -- The {\em Li\`ege Oscillations Code} is presented by \citet{rs2-apss}.

\item [$\bullet$] \NOSC\ -- The {\em Nice Oscillations Code} is presented by \citet{jp-apss}.

\item [$\bullet$] \OSCROX\ -- This English stellar oscillations code is presented by \citet{ir2-apss}.

\item [$\bullet$] \POSC\ -- The {\em Porto Oscillations Code} is presented by \citet{mm1-apss}.

\item [$\bullet$] \PULSE\ -- This Canadian stellar oscillations code is presented by \citet{sc-apss}.

\item [$\bullet$] \ROMOSC\ -- This Romanian code for {\em Linear Non-Adiabatic NonRadial Waves} is presented by \citet{ms-apss}.
  
\end{itemize}

\section{Presentation of the TASKs}\label{sec:tasks}

To perform the comparison of the participating numerical codes, the {\esta} group  has focused on specific \tasks. The first task, {\task}~1, has consisted in comparing stellar models and evolution sequences produced by eight stellar internal structure and evolution codes. For that purpose we have fixed some standard input physics and initial parameters for the stellar models to be calculated and we have defined several specific study cases corresponding to stars covering the range of masses, evolution stages and chemical compositions expected for the bulk of {\corot} target stars. The second task, {\task}~2, has consisted in testing, comparing and optimising the seismic codes by means of the comparison of the frequencies produced by nine different oscillation codes, again for specific stellar cases. Finally, the third task, {\task}~3, still in progress, is similar to {\task}~1 but for stellar models including microscopic diffusion of chemical elements. 

\subsection{\task~1: basic stellar models}

{\task}~1 has been defined with the aim of comparing the properties of stellar models and evolutionary sequences covering a rather large range of stellar parameters (initial mass, initial chemical composition and evolution stage) corresponding to the bulk of the \corot \ targets. 

Seven study cases have been considered, the specifications of which are given in Table \ref{tab:task1}. The cases are defined as follows. Seven evolutionary sequences have been calculated for different values of the stellar mass and initial chemical composition ($X, Y, Z$ where $X$, $Y$ and $Z$ are respectively the initial hydrogen, helium and metallicity in mass fraction). The masses are in the range $0.9{-}5.0$\msol. For the initial chemical composition, several pairs $(Y,Z)$ have been considered by combining two different values of $Z$ ($0.01$ and $0.02$) and two values of $Y$ ($0.26$ and $0.28$). The corresponding values of $(Z/X)$ are in the range  $0.014{-}0.029$. We have adopted the so-called GN93 solar mixture \citep[see][]{GN93} which has  $(Z/X)_\odot=0.0245$. We have therefore metallicities such that ${\rm [Fe/H]}\in [-0.25, +0.17]\ \rm{dex}$ (with the standard definition ${\rm [Fe/H]}=\log{(Z/X)}-\log{(Z/X)_\odot}$). 

\begin{table}[ht!]
\caption{Target models for {\task}~1 (see Fig.~\ref{fig:hr-task1}).
We have considered 7 cases corresponding to different initial masses, chemical compositions and evolutionary stages. One evolutionary sequence (denoted by ``Ov'' in the $\rm{5^{th}}$ column has been calculated with core overshooting -- see text).}
\centering
\label{tab:task1}
\begin{tabular}[h]{cccclr}
\hline\noalign{\smallskip}
{\bf Case} & $M/M_\odot$ & \boldmath$Y_0$ & \boldmath$Z_0$ &
  {\bf Specification} & {\bf Type} \\[3pt]
\tableheadseprule\noalign{\smallskip}
{\bf 1.1} & 0.9 & 0.28 & 0.02 &
  $X_{\mbox{\scriptsize c}}{=}0.35$ & 
  \begin{small}MS\end{small}\\[2pt]
{\bf 1.2} & 1.2 & 0.28 & 0.02 &
  $X_{\mbox{\scriptsize c}}{=}0.69$ & 
  \begin{small}ZAMS\end{small}\\[2pt]
{\bf 1.3} & 1.2 & 0.26 & 0.01 &
  $M^{\mbox{\scriptsize He}}_{\mbox{\scriptsize c}}{=}0.10$ \msol & 
  \begin{small}SGB\end{small}\\[2pt]
{\bf 1.4} & 2.0 & 0.28 & 0.02 &
  $T_{\mbox{\scriptsize c}}{=}1.9{\times}10^7$~K &
  \begin{small}PMS\end{small}\\[2pt]
{\bf 1.5} & 2.0 & 0.26 & 0.02 &
  $X_{\mbox{\scriptsize c}}{=}0.01, {\mbox{Ov}}$ & 
  \begin{small}TAMS\end{small}\\[2pt]
{\bf 1.6} & 3.0 & 0.28 & 0.01 &
  $X_{\mbox{\scriptsize c}}{=}0.69$ &
  \begin{small}ZAMS\end{small}\\[2pt]
{\bf 1.7} & 5.0 & 0.28 & 0.02 &
  $X_{\mbox{\scriptsize c}}{=}0.35$ &
  \begin{small}MS\end{small}\\[2pt]
\noalign{\smallskip}\hline
\end{tabular}
\end{table}

\begin{figure}
\centering
\resizebox{\hsize}{!}{\includegraphics{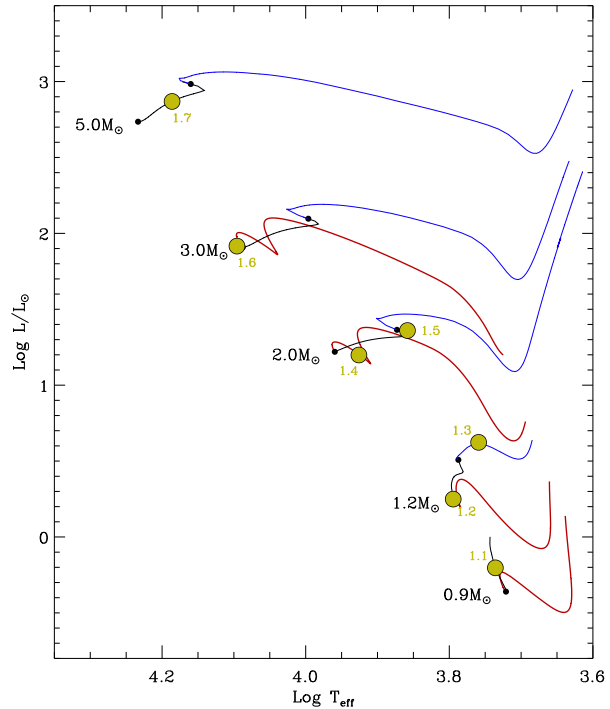}}
\caption{H--R diagram showing the targets for {\task}~1 (see Table~\ref{tab:task1}).
  Red lines correspond to the \PMS, black lines to the \MS\ and blue lines to the post-main sequence \SGB\ evolution.
  The targets are ordered in mass and age along the diagram, from Case 1.1 (bottom-right) up to Case 1.7 (top-left)}
\label{fig:hr-task1}       
\end{figure}

On each evolutionary sequence, an evolutionary stage has been selected, either on the pre-main sequence (\PMS), main sequence (\MS) or subgiant branch (\SGB). On the {\PMS} we have specified the value of the central temperature of the model ($T_{\mathrm c}=1.9 \times 10^7\ {\mathrm K}$). On the {\MS}, we have fixed the value of the central hydrogen content: $X_{\mathrm c}=0.69$ for the model close to the zero age main sequence (\ZAMS), $X_{\mathrm c}=0.35$ for the model in the middle of the {\MS} and $X_{\mathrm c}=0.01$ for the model close to the terminal age main sequence (\TAMS). On the \SGB, a model can be chosen by specifying the value of the mass $M_{\mathrm c}^{\mathrm{He}}$ of the central region of the star where the hydrogen abundance is such that $X\leq 0.01$. We chose  $M_{\mathrm c}^{\mathrm{He}}=0.10$ \msol. Figure \ref{fig:hr-task1} displays the location in the H--R diagram of the targets for {\task}~1 together with their parent evolutionary track.

All models calculated for {\task}~1 are based on rather simple input physics, currently implemented in stellar evolution codes and  one model has been calculated with overshooting (see Sect. \ref{sec:physics} below).

The first results of {\task}~1 have been presented by \cite{2006corm.conf..363M}. Detailed results are presented in this volume \citep{mm2-apss,yl3-apss,jm-apss,mam-apss}.

\subsection{\task~2: oscillation frequencies}

Under \task~2 we have aimed at evaluating the precision and uncertainties affecting the determination of oscillation frequencies.
In particular we have studied the numerical precision of the frequencies using different meshes and with a wide range of options to solve the equations for linear adiabatic (radial and non-radial) oscillations.
In particular two steps have been considered in this \task:
\begin{itemize}
\item [] Step 1 -- comparison of the frequencies from different codes for the same stellar model;
\item [] Step 2 -- comparison of the frequencies from the same seismic code for different stellar models of the same stellar case.
\end{itemize}
In this exercise stellar models at the middle of the main sequence, produced by \astec\ and \cesam, have been used with a mass of $1.5~M_\odot$.
The models have been provided with different number of mesh points.

Preliminary results of this task were reported by \cite{am-eas} while the final results of Step 1 are discussed in this volume by \cite{moya-apss}. Step 2 remains to be done.

\subsection{\task~3: stellar models including microscopic diffusion}

The goals of {\task}~3 are to test, compare and optimise stellar evolution codes which include microscopic diffusion of chemical elements. At this stage we only consider diffusion resulting from pressure, temperature and concentration gradients \citep[see][]{tm07} while we do not take into account diffusion due to the radiative forces, nor the extra-mixing of chemical elements due to differential rotation or internal gravity waves \citep[see][]{alecian07,mathis07,zahn07}. The other physical assumptions proposed as the reference for the comparisons are the same as used for {\task}~1 and no overshooting (see Sect. \ref{sec:physics}).

Three study cases have been considered for the models to be compared. Each case corresponds to a given value of the stellar mass (see Table \ref{tab:task3}).  We chose rather low values of the masses (i.e. $M<1.4$\ \msol) in order to keep in a mass range where radiative accelerations can be omitted. Furthermore, this avoids the problems occurring at higher masses where the use of microscopic diffusion alone produces a very substantial depletion of helium and heavy elements at the surface (and a concomitant increase of the hydrogen content) and in turn requires to invoke other mixing processes to control the gravitational settling \citep[see, for instance][]{1998ApJ...504..539T}. 

\begin{table}[ht!]
\caption{Target models for {\task}~3. {\em Left}: Three cases with corresponding masses and initial chemical composition. {\em Right}: Three evolutionary stages examined for each case. Stages A and B are respectively in the middle and end of the MS stage. Stage C is on the SGB.}
\centering
\label{tab:task3}       
\begin{tabular}{cccc}
\hline\noalign{\smallskip}
\begin{small}{\bf Case}\end{small} & $\! M/M_\odot \!$ & \boldmath$Y_0$ & \boldmath$Z_0$ \\[3pt]
\tableheadseprule\noalign{\smallskip}
\bfseries{3.1}& $1.0$ & $0.27$ & $0.017$ \\
\bfseries{3.2}& $1.2$ & $0.27$ & $0.017$ \\
\bfseries{3.3}& $1.3$ & $0.27$ & $0.017$ \\
\noalign{\smallskip}\hline
\end{tabular}
\hfill
\begin{tabular}{ccc}
\hline\noalign{\smallskip}
\begin{small}{\bf Stage}\end{small} & \boldmath$X_{\rm c}$ & \boldmath$M^{\mbox{\scriptsize He}}_{\mbox{\scriptsize c}}$ \\[3pt]
\tableheadseprule\noalign{\smallskip}
\bfseries{A}& $0.35$ & - \\
\bfseries{B}& $0.01$ & - \\
\bfseries{C}& $0.00$ & $0.05\ M_{\star}$ \\
\noalign{\smallskip}\hline
\end{tabular}
\end{table}

\begin{figure}
\centering
\resizebox{\hsize}{!}{\includegraphics{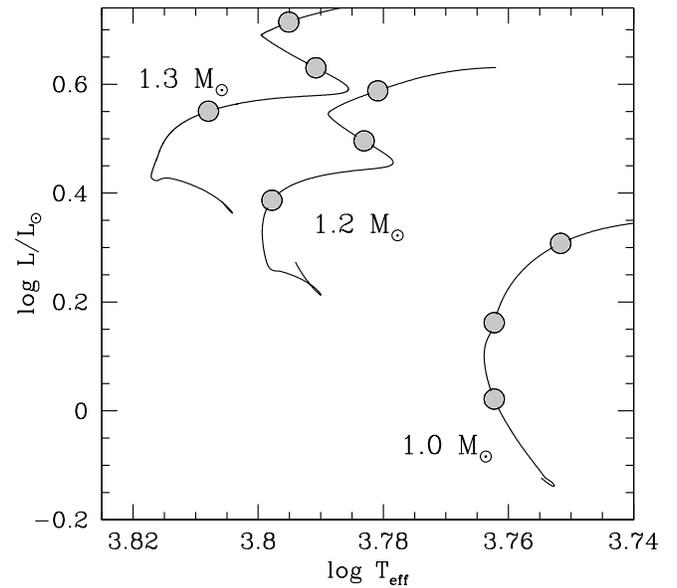}}
\caption{H--R diagram showing the targets for {\task}~3 (see Table~\ref{tab:task3}). 
Evolutionary tracks correspond to Case 3.1 (bottom-right), Case 3.2 (middle) and Case 3.3 (top-left). On each track  filled circles indicate the stages A, B, C.
  }
\label{fig:hr-task3}       
\end{figure}

For the three cases we have adopted a chemical composition close to the solar one ($Z/X=0.0243$).
For each case, models at different evolutionary stages have been considered. We focused on three particular evolution stages: middle of the \MS, {\TAMS} and {\SGB} (respectively stage A, B and C). Figure \ref{fig:hr-task3} displays the location in the H--R diagram of the targets for {\task}~3 together with their parent evolutionary track.

Preliminary results of {\task}~3 have been presented by \citet{yl-eas,jm-eas,jcd-eas}. Advanced comparisons  are  presented in this volume \citep{yl3-apss}.

\section{Specifications for the \tasks}\label{sec:physics}

In order to reduce the sources of discrepancies in the comparison due to sources not relevant in this context it has been necessary to specify the initial parameters and the physics to be used in the models.
Special care has also been given to the exchange of data in order to facilitate the exchange of models and evolutionary tracks and their use in detailed comparisons.

Here we briefly describe the common specifications used for all \tasks. Further details can be found in the documentation available at the \esta\ web-site\footnote{at {\tt\small http://www.astro.up.pt/corot/compmod/task1}}.

\subsection{Initial parameters}

A list of reference values of some astronomical and physical constants have been specified to safeguard consistency in the comparison of the output from different codes. We have also fixed the mixture of heavy elements to be used in the calculations.

\begin{itemize}

\item [a)] {\em Physical constants:} 
We have specified the values of the physical constants necessary for the calculation of a stellar model according to \cite{1987RvMP...59.1121C} and \cite{1994chcp.book.....L} (see Table~\ref{tab:const}). Furthermore a reference set of values is provided for the mass of neutron and atomic mass of the elements from \elem{H}[1] to \elem{O}[17].

\begin{table*}[ht!]
\caption{Some of the physical constants necessary for the calculation of a stellar model. The units are {\it cgs}, except where otherwise noted.}
\centering
\label{tab:constants}
\begin{tabular}[h]{rrl}
\tableheadseprule\noalign{\smallskip}
Boltzmann's constant & $k=$ & $1.380658\times 10^{-16}$\\[2pt]
Atomic mass unit & $m_u=$ &  $1.6605402 \times 10^{-24}$\\[2pt]
Perfect gas constant & $\mathcal{R}=$  & $8.3145111\times 10^{7}$\\[2pt]
Electron mass & $m_e=$ & $9.1093897\times10^{-28}$\\[2pt]
Electron charge & $e=$ &  $1.602177333\times 10^{-19}~C$\\[2pt]
Planck's constant & $h=$ &  $6.6260755\times 10^{-27}$\\[2pt]
Speed of light & $c=$ &  $2.99792458\times10^{10}$\\[2pt]
Radiation density constant & $a=$ &  $7.5659122\times 10^{-15}$\\[2pt]
Stefan-Boltzmann constant & $\sigma=$ &  $5.67051 \times 10^{-5}$\\[2pt]
Electron--Volt & 1~eV~= &  $1.60217733\times 10^{-11}$\\[2pt]
Atomic weight of hydrogen & $A_H=$ &  1.00782500 \\[2pt]
Atomic weight of helium & $A_{He}=$ &  4.00260330 \\[2pt]
Ionisation potential for H & $\chi_H=$ &  13.595~eV \\[2pt]
1$^{\rm st}$ ionisation potential for He & $\chi_{He}=$ &  24.580~eV \\[2pt]
2$^{\rm nd}$ ionisation potential for He & $\chi_{He^+}=$ &  54.403~eV \\[2pt]
\noalign{\smallskip}\hline
\label{tab:const}
\end{tabular}
\end{table*}

\item [b)] {\em Astronomical constants}:
We have chosen the values to consider for the global parameters of the present Sun. We have adopted 
$$R_\odot = 6.9599\times  10^{10}\ {\rm cm},$$
for the solar radius \citep{1973asqu.book.....A},
$$L_\odot=3.846\times 10^{33}\  {\rm{erg~s^{-1}}},$$
for the solar luminosity \citep{1986Sci...234.1114W} and
$$M_\odot=1.98919\times 10^{33}\ {\rm g},$$
for the solar mass.
Here {\msol} is obtained from $(GM_\odot)=1.32712438\times 10^{26}\ {\rm cm^{-3}s^{-2}}$ \citep{1994chcp.book.....L} and 
$$G=6.6716823\times 10^{-8} {\rm cm^{-3}g^{-1}s^{-2}}.$$

The value {\rsol} refers to the radius of the model layer where $T=T_{\rm eff}=5777.54$\ K.

\item [c)] {\em Initial abundances of the elements and heavy elements mixture:}
All models are calculated with the classical GN93 solar mixture of heavy elements \citep{GN93}. As shown in Tables \ref{tab:task1} and \ref{tab:task3} the mass fractions of hydrogen ($X$), helium ($Y$) and heavy elements ($Z$) are specified for each model.

\end{itemize}

\subsection{Input physics}

In order to be able to make basic comparisons between the codes, we chose some reference input physics that have the advantage to be currently implemented in stellar evolution codes. 

\begin{itemize}

\item [a)] {\em Equation of State:}
We chose to use the OPAL equation of state (EOS) in its 2001 version \citep{2002ApJ...576.1064R} available at the OPAL Web site\footnote{{\tt\small http://www-phys.llnl.gov/Research/OPAL}}. It consists in a series of tables and in an associated interpolation package.
The tables provide a set of thermodynamic quantities, i.e. pressure, internal energy, entropy, specific heat $C_{\rm v}$, $\chi_\rho=(d\ln P/d \ln \rho)_T$, $\chi_{\rm T}=(d\ln P/d \ln T)_{\rho}$ and the adiabatic indices $\Gamma_1$, $\Gamma_2/(\Gamma_2{-}1)$ and $(\Gamma_3{-}1)$, at given values of the  temperature, density, $X$ and $Z$. However, there are differences in the handling of the tables by the different codes. Some use the interpolation package provided by the OPAL group, others have developed their own interpolation package. In addition, it has been pointed out by \cite{2003ApJ...583.1004B} that there are some inconsistencies in the OPAL tables with the result that the tabulated values of the adiabatic indexes $\Gamma_1$, $\Gamma_2/(\Gamma_2{-}1)$ and $(\Gamma_3{-}1)$ may differ by several per cent from the values that can be recalculated from the tabulated values of $P$, $C_{\rm v}$, $\chi_\rho$ and $\chi_{\rm T}$ by means of thermodynamic relations.
Furthermore, in the course of the comparisons undertaken by the {\esta} group it has been shown by one of us (I.W. Roxburgh) that the tabulated values of the specific heat $C_{\rm v}$ in OPAL tables are incorrect, and also acknowledged by the OPAL team that  this parameter should better be obtained from the other  quantities provided by
OPAL.
 For these reasons, some codes only draw $P$, $\chi_\rho$, $\chi_{\rm T}$ and $\Gamma_1$ from the OPAL EOS tables and apply thermodynamic relations to derive the other thermodynamic quantities.

\item [b)] {\em Opacities}:
We chose to use the 1995 OPAL opacity tables \citep{ir96} complemented at low temperatures by the \cite{af94} tables, and not to include the conductive opacity. However slightly different OPAL tables are actually used by the different codes. Pre-calculated OPAL tables have been made available by the OPAL group either prior to publication or later on the OPAL Web site. {They have been obtained for a ``reduced'' GN93 solar mixture, 
i.e. while the GN93 mixture is given for 23 elements, the OPAL opacity calculation only considers 19 elements but adds the abundances of ``lacking" elements to the abundances of nearby elements (having close atomic number).}
 Also, on the OPAL Web site there is the option to calculate tables on-line for any desired 19 elements-mixture and to pass or not a smoothing filter to reduce the random numerical errors that affect the OPAL opacity computation.
Different groups used either one of these possibilities. Moreover, some codes use the interpolation routines provided by the OPAL group, others designed their own routines.

\item [c)] {\em Nuclear reaction rates:}
No specifications have been given for the choice of the nuclear network. The basic pp chain and CNO cycle reaction networks up to the \elem{O}~[17](p,$\alpha$)\elem{N}[14] reaction have generally been used. Some models have been calculated with the option/assumption that either one or all the light elements (\elem{He}[3], \elem{Li}[7], \elem{Be}[7], \elem{H}[2]) are at equilibrium while other models have followed them explicitly.  We chose to compute the nuclear reaction rates from the analytical formulae provided by the NACRE compilation \citep{1999NuPhA.656....3A}. Different prescriptions have been used for the screening. Most often, weak screening has been assumed under the \cite{1954AuJPh...7..373S} formulation. In that case the screening factor is written
$$f=\exp\left(A z_1 z_2 \sqrt{\frac{\rho \xi}{{T}^3}}\right),$$
where $z_1$ and $z_2$ are the charges of the interacting nuclei. Some codes have used the expression (4-221) of \cite{1968psen.book.....C} giving $A=1.88\times 10^8$ and $\xi=\sum_{i} z_i(1{+}z_i)x_i$ where $x_i$ is the abundance per mole of element $i$.  Finally, in the nuclear reaction network the initial abundance of each chemical species is split between its isotopes according to the isotopic ratio of nuclides for which values have been given (see the ESTA Web site\footnote{at {\tt\small http://www.astro.up.pt/corot/compmod/task1}}). 

\item [d)] {\em Convection and overshooting:}
We chose to use the classical mixing length treatment of  \cite{1958ZA.....46..108B} under the formulation of \cite{1965ApJ...142..841H} taking into account the optical thickness of the convective bubble. The value of the mixing length parameter has been chosen to be $\alpha_{\rm MLT}=1.6$. The onset of convection is determined according to the Schwarzschild criterion $(\nabla_{\rm ad}{-}\nabla_{\rm rad}{<}0)$ where $\nabla_{\rm ad}$ and $\nabla_{\rm rad}$ are respectively the adiabatic and radiative temperature gradient. In models with overshooting, the convective core is extended on a distance $l_{\rm ov}=\alpha_{\rm ov}\times \min(H_p, R_{\rm cc})$ where $H_{p}$ is the pressure scale height and $R_{\rm cc}$ the radius of the convective core. We have chosen the value of the overshooting parameter to be $\alpha_{\rm ov}=0.15$. The core is mixed in the region corresponding to the convective and overshooting region. In the overshooting region the temperature gradient is taken to be equal to the adiabatic gradient. Note that some codes have provided models including overshooting but following different parametrizations (e.g. {\STAROX} takes the temperature gradient to be equal to the radiative gradient in the overshooting region while {\FRANEC} uses a formulation of overshooting depending on the mass). 

\item [e)] {\em Atmosphere:}
Eddington's grey $T(\tau)$ law has been prescribed for the atmosphere calculation:
$$T = T_{\rm eff} \left[\frac{3}{4}\left(\tau+\frac{2}{3}\right)\right]^\frac{1}{4},$$
where $\tau$ is the optical depth. The radius of the star is taken to be the bolometric radius, i.e. the radius at the level where the local temperature equals the effective temperature ($\tau=2/3$ for the Eddington's law).  Codes manage differently the integration of the hydrostatic equation in the atmosphere from a specified upper value of the optical depth $\tau$ to the connexion with the envelope.

\item [f)] {\em Microscopic diffusion:}
We only considered the diffusion of helium and heavy elements due to pressure, temperature and concentration gradients and we neglected radiative accelerations that are not yet fully included and tested in the participating codes. We did not impose the formalism to be used. As reviewed by \citet{tm07}, two approaches to obtain the diffusion equation from the Boltzmann equation for binary or multiple gas mixtures can be followed: one is based on the Chapman-Enskog theory \citep[][hereafter CC70]{1970mtnu.book.....C} and the other on the resolution of the Burgers equations \citep[][hereafter B69]{B69}. In both methods, approximations have to be made to derive the various coefficients entering the diffusion equations, in particular the diffusion velocities which are written as a function of the collision integrals.  In the stellar evolution codes which have participated in {\task}~3, different treatments of the diffusion processes have been adopted. The {\tgec} code follows the CC70 approach. The other codes follow the B69 approach but differently: (i) the {\astec} and {\franec} codes use the simplified formalism of \citet[][hereafter {MP93}]{MP93}, (ii) the {\cles} and {\garstec} codes compute the diffusion coefficients by solving Burgers' equations according to the formalism of \cite{TBL94} and (iii) {\cesam} provides two options to compute diffusion velocities, one  based on the {MP69} approximation, the other on Burger's formalism, with collision integrals derived from \citet{1986ApJS...61..177P}. Furthermore, the number of chemical elements considered in the diffusion process differs in the different codes and some codes consider that all the elements are completely ionised while others calculate the ionisation explicitly.


\end{itemize}

\subsection{Output format}


Each code had previously designed its own output format.
But to facilitate the comparisons we have adapted some of the output in order to include all information required for detailed code-to-code comparisons of models.
In most exercises we have adopted the file format prescribed by the {\small GONG} solar model team which is described in the documentation files in the \ADIPLS\ package and on the {\esta} Web site\footnote{at {\tt\small http://www.astro.up.pt/corot/ntools/}}. This permits a rather easy treatment of the models which can be read, plotted and compared by means of the programme tools available in the \ADIPLS\ package. In addition, a tool {\small MODCONV} and associated documentation have been made available on the {\esta} Web-site by M.~Monteiro to allow the direct conversion between the different formats available.

In all cases {\esta} participants have been asked to provide {\small ASCII} files ready for comparisons which contain, for the {\task} case considered, the global properties of the model (name, mass, age, initial composition, luminosity, photospheric radius, etc.) as well as a wide range of model variables at each mesh point (distance $r$ to the centre, mass inside the sphere of radius $r$, pressure, density, temperature, chemical composition, opacity and several other physical parameters including quantities of interest for the computation of adiabatic oscillations like the 
Brunt-V\"ais\"al\"a frequency). 

Modelers have also provided {\small ASCII} files giving the variation with time of some global or internal properties of their evolution sequences (luminosity, effective temperature, radius, central hydrogen content, surface helium and heavy-element abundances, radius and mass of the convective core and depth and mass of the convection envelope, etc).

\subsection{Oscillations}

Stellar oscillations modelers involved in \task~2 were asked to provide
adiabatic frequencies in the range $[20,2500] \,\mu$Hz and with spherical
degrees $\ell$=0,1,2 and 3. To obtain the solution of the equations, the modelers were asked to
adopt the following specifications:

\begin{itemize}
\item [a)] {\em Mesh:} Use the mesh provided by the equilibrium model (no re-meshing).
\item [b)]  {\em Outer mechanical boundary condition:} Set the Lagrangian perturbation to the pressure to zero ($\delta
 P=0$).
\item [c)]  {\em Physical constants:} Use the same as in \task~1 and 3.
\item [d)] {\em Equations:} Use linear adiabatic equations.
\end{itemize}

However, different methods and assumptions are made to write and to solve numerically the differential equations in the participating oscillation codes. In particular:

\begin{itemize}

\item [a)] {\em Set of eigenfunctions:} Either the Lagrangian or the Eulerian
 perturbation to the pressure ($\delta P$ or $P^\prime$) have been used which affects the form of the equations.

\item[b)] {\em Order of the integration scheme:} Either a second-order or a fourth-order scheme has been used.

\item[c)] {\em Richardson extrapolation:} Some codes using a second-order
 scheme have the possibility to use Richardson extrapolation
 \citep{shiba} to decrease the truncation error. The combination of a $2^{\rm nd}$ order scheme with Richardson extrapolation leads to errors which scale as ${\cal N}^{-4}$, where ${\cal N}$ is the number of mesh points
 \citep[e.g.][]{Christ1994}.

\item[d)] {\em  Integration variable:} Either the radius $r$ or the ratio $r/P$ have been used as integration variables.

\item[e)] {\em Physical constants:} Some codes use a value of the gravitational constant $G$ slightly different than the one specified in the other \tasks. As pointed out by \citet{moya-apss}, using different values of $G$ in the stellar structure equations and in the 
 oscillation equations gives rise to inconsistencies.

\end{itemize}

The consequences of these choices are discussed in details in \citet{moya-apss}.


\section{Reference grids}\label{sec:grids}

In parallel to the three \tasks, grids of models have been especially calculated with the {\CESAM} and {\CLES} codes for masses in the range $0.8{-}10~M_\odot$ and chemical compositions $\mbox{[Fe/H]}{=}0.0$ and $-0.10$.  These reference grids have been used to locate {\corot} potential targets in the H--R diagram in the process of target selection (see Figure~\ref{fig:targets}) and to study some $\delta$ Scuti candidates for {\corot} \citep{2005AJ....129.2461P}. In addition, associated oscillation frequencies have been calculated for selected models in the grid either with the {\ADIPLS}, {\POSC} or {\LOSC} codes. This material is described in this volume by \citet{jm1-apss,yl2-apss,mmf-apss}.

In addition, a database of models of $\beta$~Cephei stars and their oscillation frequencies called {\it BetaDat} has been designed in the context of {\esta} \citep{at-apss}, although the models do not use exactly the same input physics and constants. It is accessible through a Web interface\footnote{{\tt\small http://astrotheor3.astro.ulg.ac.be/}}.

\section{Conclusion}

In this paper we briefly presented the different activities undertaken by the evolution and seismic tools activity \citep[\esta, see][]{2006corm.conf..363M} organised under the responsibilities of the {\em Seismology Working Group} \citep[see][]{michel06} of the \corot\ mission.
A description of the \esta\ team and \tasks\ aimed at improving stellar evolution codes and oscillation codes is presented.
This provides the context and the reference for the more detailed works published in this volume with the descriptions of the codes and the results of the comparisons.

The seismic study of the Sun (helioseismology) had already pushed the need to calculate solar models to a new level of accuracy due to the necessity to calculate theoretical oscillation frequencies that could match the observed solar values.
As we now approach the same level of precision for other stars in asteroseismology the codes used to model the evolution of stars of different masses have now to reach the same high precision for stellar regimes very different from the Sun.
As the physics dominating different regimes in the H--R diagram are different a strong effort towards this is required.

In the work developed by ESTA we are pursuing that goal. The results of this exercise so far have shown that we are able to meet the present requirements.
But more work is planned to address some pending aspects of the physics that must be included in the models with sufficient precision to be able to reproduce the observed seismic behaviour.
This is an open project as the arrival of data from \corot\ and other future projects will required our tools to improve further the precision of the models in order to test the highly accurate data made available for asteroseismology.

\begin{acknowledgements}
This work is being supported in part by the
European Helio- and Asteroseismology
Network (HELAS), a major international collaboration
funded by the European Commission's Sixth Framework Programme.
MJPFGM is supported in part by FCT and FEDER (POCI2010) through projects
{\small POCI/CTE-AST/57610/2004} and {\small POCI/V.5/B0094/2005}. A.M. acknowledge financial support of the Spanish PNE under Project number ESP 2004-03855-C03-C01. J.M. has received a financial support from the Federal Science Policy Office (BELSPO) in the frame of the ESA/Prodex program, contract C90199 -- CoRoT -- Preparation to exploitation.
\end{acknowledgements}





\end{document}